# The Relationship between Digital RMB and Digital Economy in China

Chang Su[a#*], Wenbo Lyu[b#], Yueting Liu[b]

[a] *Judge Business School, University of Cambridge, CB2 1AG, UK*
[b] *Nanyang Technology Entrepreneurship Center, Nanyang Technological University, Singapore, 639798, Singapore*
* Corresponding author. Tel/Fax: +44 0 7456230159; E-mail: cs2134@cam.ac.uk
# C.S. and W.L. contributed equally to this work

**Abstract:** By comparing the historical patterns of currency development, this paper pointed out the inevitability of development of digital currency and the relationship between the digital currency and the digital economy. With the example of China, this paper predicts the future development trend of digital currency. In the context of the rapid development of private cryptocurrency, China launched the digital currency based on serving the digital economy and committed to the globalisation of the digital renminbi (RMB) and the globalisation of the digital economy. The global economy in 2022 ushered in a stagnation, and China treats digital fiat currency and the digital economy development as a breakthrough to pursue economic transformation and new growth. It has become one of the forefront countries with numerous experiences that can be learned by countries around the world.

## 1. Chinese Currency in The History of Money

The history of the development of money shows that money has gradually evolved from physical money to credit money. In the early days, physical money acted as a medium of exchange and performed the monetary function of a measure of value by virtue of its scarcity, portability and storage resistance, which improved the efficiency of economic activities and freed it from the restriction of the single transaction mode of barter. With the advancement of smelting technology and the establishment of state power, money completed the transformation from physical money to metallic money, and shifted from full-value metallic money to under-value metallic money. The credit attributes of legal tender were initially developed at this point. However, the actual supply of metallic money was constrained by the mining and refining technology and scale, which was difficult to match the growth of tradable social wealth [1], easily causing deflation and "fire consumption" speculation, and even triggering monetary imbalance and war between different countries.

Along with the advancement of paper-making technology and the increase of national sovereignty, paper money secured by sovereign credit was created. Money was separated from metal value and transformed into a reference for social wealth, becoming a pure value symbol to facilitate sovereign states to issue legal tender based on the total amount of tradable wealth in society. Driven by the dual factors of credit money system and several technological revolutions leading to productivity progress, the total amount of social assets and the total wealth of society denominated in money increased by leaps and bounds. Even so, in terms of anti-counterfeiting and durability, credit banknotes also require a lot of resources and technologies to process and modify the paper, resulting in a certain manufacturing cost. With a huge number of bases, this manufacturing cost becomes very large. Small coin minting, on the other hand, requires materials such as aluminum-magnesium alloy, stainless steel, nickel-plated steel cores, and copper-plated steel cores, which are even more costly than paper money

In addition, the cost of transporting and storing physical cash in commercial banks and other institutions is also quite high. According to the Fed, the U.S. monetary business budget in 2020 is as high as $880 million, of which the proposed budget for fixed printing costs is about $500 million. Due to the high cost and the consideration of banknote carrying and anti-theft and anti-counterfeiting, the process of making electronic money and digital money has been accelerated. Today, electronic money has been developed by leaps and bounds. People can use various payment method, such as PayPal, Apple Pay, Alipay and so on to make transaction. Cryptocurrencies such as Bitcoin and Libra have also been widely circulated. China is the first country to launch digital legal currency, and it is also the only country that has experienced the evolution history of full currency from physical money, metallic money, to e-money and digital money.

Electronic currency and digital currency have strong similarities, but it should be noted that digital currency is a currency created based on digital technology, and can realize the function of currency in a specific digital field, while electronic currency usually refers to the electronic presentation of legal currency, such as the display of currency balances and the electronic operation form of transfer and collection in the mobile banking of commercial banks. To sum up, electronic currency is essentially an electronic presentation of legal tender, while digital money is essentially a digital program code that is recognized and used in a specific domain category.

**2. Status Quo of Global Digital Legal Currency**

According to the Bank for International Settlements (BIS) data, 62% of central banks have promoted digital currency into the empirical stage in 2020, an increase of 20 percentage points from 2019. Although different countries are looking at digital fiat currencies in different directions, some focusing on the wholesale type for large trade and some on the retail type for small-value consumption, the sovereign logic of digital fiat currencies is the same, they are all designed to counter digital currencies issued by private sector or digital fiat currencies of other sovereign countries. When a country's sovereignty loses control over the currency in circulation (including digital currency), the country's sovereignty will be seriously challenged, which is unacceptable and intolerable at the state power level.

At present, there are three main attitudes towards digital currency in the world: active promotion in countries such as China and Europe, wait-and-see review in the United States, and firm rejection by a few countries. Although China actively promotes digital currency, it only recognizes digital RMB, and all other private digital currencies are not legally recognized. The United States attitude is also evolving under scrutiny, and the concept of digital dollar was proposed in early 2022 in a change of pace, reflecting the rich theoretical experience the U.S. government has accumulated in the development of digital dollar from the wild exploration of private digital currency. Financial authorities in Europe and the United States, as well as Japan and Singapore, have embarked on the exploration of joint digital currencies, demonstrating that these countries are not only exploring the technical means of digital fiat money, but are also attempting to achieve a barrier-free cross-border circulation of international digital fiat money, or even a digital fiat money that is recognized worldwide. Seen from the study *Central bank digital currencies: foundational principles and core features* released by the Bank for International Settlements on October 9, 2020, which was jointly researched and completed by the Bank of Canada, European Central Bank, Bank of Japan, Sveriges Riksbank, Swiss National Bank, Bank of England and the Board of Governors Federal Reserve System, the intentions of the world digital fiat currency alliance are beginning to emerge.

Although capitalist and socialist countries have different ideologies, they both seek a digital legal currency solution for mutual integration in the economic field. The future world will develop into a two-level or even multi-level digital currency structure. Perhaps in the infinitely distant future, the final competition outcome of the digital legal currency of various countries may be end-up as a digital super currency, that would be calculated by a package of digital currencies of countries in the world with reference to the national strength, which is the most scientific and fair digital currency that dynamically adjusts and changes with the contribution rate of each country to the world economic development.

In the current global financial environment, the euro shows increasing potential to overtake the dominant currency status of the US dollar. As the member of EU expands further, the popularity of the euro continues to catch up with the U.S. dollar [2]. Russia-Ukraine war in 2022 makes the relationship between Europe and the United States changes from Trump's tariff confrontation to Biden's full solidarity, providing a good political background for the euro to expand its international influence, and laying the groundwork for the future to surpass the US dollar.

Significantly different from Europe and the United States, China's central bank

established a digital legal currency research group as early as 2014. Today, the digital RMB is relatively mature, and has been promoted in three pilots, expanding its application scenarios and scope. In December 2020, the Central Bank Digital Research Institute and UnionPay launched business cooperation to jointly study innovative applications in online and offline payment scenarios and other fields, solve technical problems such as double-spending, and realise the application scenarios of digital RMB for daily bill payment, food delivery service, shopping and consumption, and e-government, etc.

The People's Bank of China (PBoC) has released a digital version of legal tender known as Digital RMB, which is also called Digital Currency Electronic Payment (DCEP). It is traded for the public by designated entities such as commercial banks. It supports loosely-coupled account linkage and offline payment functions based on a generic account structure. It is equivalent to banknotes and coins, and introduces some innovative features of blockchain and smart contracts to support controlled anonymity. The DCEP implements the "one currency, two repositories, and three centres" operating structure and the "central bank & commercial bank" dual issuance model. It is positioned as the M0, and its essence is to act as and replace part of the cash in circulation. The design of the use of digital RMB is also delicate. The requirements for the small value consumption demand side are not high, and even people without Chinese nationality can apply for and use digital RMB. From the perspective of large-value transaction, company verification and personal verification are required for users, and different rule requirements and conveniences are given for different flow amounts.

The DCEP is built on blockchain technology and overcomes blockchain's typical disadvantages. The stored transaction information on the blockchain is publicly distributed, and the blockchain information is continuously added as transactions occur, constituting an increasingly long chain of digital codes, overlaid with mathematical function formulas for encryption, which is currently considered one of the most secure technologies. However, it is unable to support high-speed and high-frequency mass transactions due to the possibility of privacy leakage by making transaction information public, as well as the distributed and large number of calculations to be performed for each encryption. These are the problems that digital RMB needs to solve urgently. Under the unified top-level design rules, combined with the advantages of blockchain, the central bank adopts a centralized computing method to achieve controllable anonymity and high-speed computing with innovative technologies such as asymmetric encryption, traceability, and smart contracts coupled with blockchain technology, which perfectly solves the above problems. It lays the technical foundation for the expansion of DCEP [4]. For example, the digital RMB retains the centralized technical characteristics, which can greatly improve the transaction efficiency, reaching a processing capacity of 300,000 transactions per second, while the decentralized Bitcoin Alliance platform can only process about 7 transactions per second, which cannot meet the demand for high frequency transactions.

## 3. Liquidity Analysis of Digital RMB to The Money Market

The meaning of money market liquidity is relatively broad and divided into three dimensions:
(1) Narrow liquidity, which reflects the amount of available funds in the banking system, represents the liquidity situation in the overall money market and is the starting point of liquidity transmission.
(2) Broad liquidity, which reflects the liquidity situation of the corporate/personal/non-bank entities, represents the liquidity of the entire real economy.
(3) Financial liquidity, reflecting the situation of the funding surface used to allocate financial assets such as stocks and bonds.

Financial liquidity is the most special and is usually characterized by asymmetric flows in both directions. For example, after the vast majority of retail investors convert their cash into equity assets, whether at a profit or a loss, their stock market funds are realized as M0, which is usually not used for consumption, but for reinvestment or long-term savings, becoming a positive M0 to M2 differential. The funds in the stock market do not achieve monetary multiplier effect so the growth rate of M0 is the source of the stock market's rise. Without considering the time point when financing and dividends are paid in the stock market, the speculative transactions in the stock market take up cash that could have been involved in the monetary multiplier amplification impact of the development of the real economy. Excessive speculation in the stock market is very detrimental to the development of the real economy. Financial asset market liquidity has thus become an area of regulatory focus.

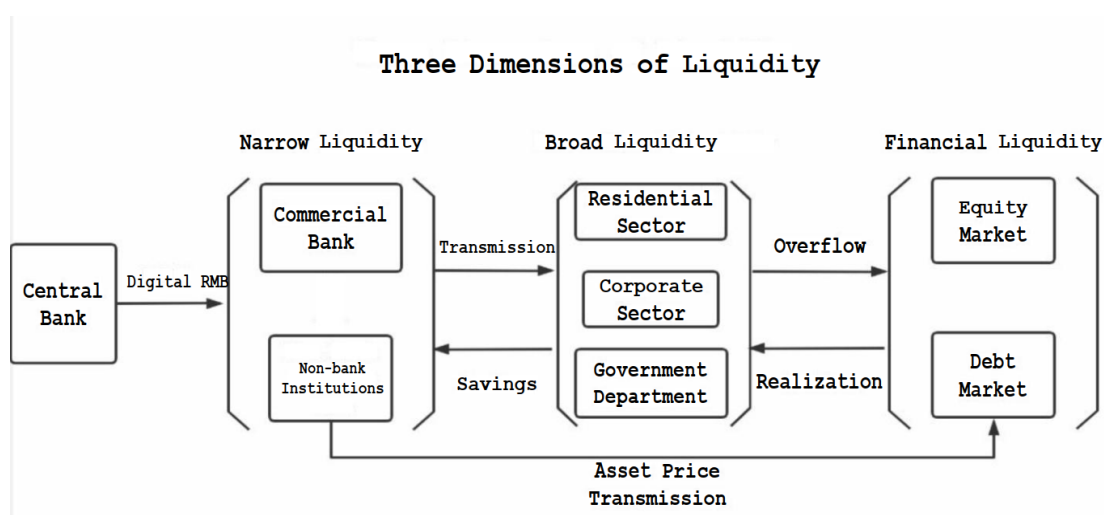

Figure 1: Three Dimensional of Digital RMB's Liquidity

The DCEP is distributed to the market through the "central bank-commercial bank" channel and can be accurately issued and tracked through technical means, which plays a role of precise execution and real-time feedback on the central bank's monetary policy, financial policy, and fiscal policy. Moreover, DCEP gives strong technical support in all three dimensions of liquidity, both to adapt to changes in monetary policy and to achieve precision in the transmission of the value of monetary assets. These are unattainable with paper money.

The DCEP, a macro-control tool, is of great significance. Based on digital technology and big data, the central bank can track and monitor all digital RMB issued in real time, count the total amount and structure of currency in real time, and provide data support for the formulation of monetary policy. At the same time, through the use of blockchain smart contract technology, the central bank can set effective conditions such as time, quantity, and investment direction in advance in the process of placing DCEP, so as to achieve traceability, programmability, and even set specific effective time-limited rules to enable monetary policy regulation more direct, precise and effective. The flow of digital currency can also be known throughout the process, and accurate statistics can be realized for the value chain of the whole industry, which can be used as the basis for the further adjustment of monetary and fiscal policies. Reserve requirements, the discount rate, and open market operations are tools of traditional monetary policy instruments, which influence the pace of macroeconomic development by adjusting the market demand for money. However, the statistics of market currencies often have a certain error rate, and DCEP can provide more accurate data, avoiding the problems of "a shift of focus from the real economy to the virtual economy " and potential resistance to counter-cyclical regulation. [5]

From a microscopic point of view, the DCEP will bring technological innovation to the identity authentication methods of commercial banks and payment institutions, thereby improving payment efficiency and operational efficiency, reducing currency manufacturing costs and use costs, and optimizing anti-money laundering systems. Furthermore, it will also have a positive impact on commercial banks' payment and settlement, business processing, operating costs, compliance and service quality, facilitating the realization of inclusive financing, lowering social financing costs and thresholds, reducing the risk of bad and doubtful debts, and greatly enhancing the financial environment in China's digital economy. These advantages and experiences can be directly extended and replicated in the implementation of digital legal tender in other countries.

4. RMB Globalisation and Digital RMB Globalisation

PBoC has established RMB clearing banks in over 25 countries and regions outside of mainland China. Currently, RMB accounts for 10.92% of the IMF SDR. In 2019, the cross-border RMB transaction volume reached 19.67 trillion yuan, and this amount increased to 28.39 trillion yuan in 2020, a year-on-year increase of 44.3%. In the first half of the year 2021, RMB cross-border receipts and payments amounted to CNY 17.57 trillion, up 2.4 percentage points from the same period last year. Although the internationalisation of the RMB has gradually accelerated and RMB has become the world's fifth largest currency year-round, it is highly disproportionate to the actual size of China's economy and the scale of its foreign trade and investment. This is directly related to the fact that the current cross-border payment transfer and clearing are highly dependent on the SWIFT system and the Clearing House Interbank Payment System (CHIPS) under the US dollar system. Combined with China's currency clearing system CIPS, the digital RMB is likely to become an alternative to cross-border payments, in pursuit of cross-border transaction currency status that match the comprehensive national strength and foreign trade levels.

At present, events such as Russia being kicked out of the SWIFT system and Saudi Arabia proposing the yuan-priced oil contracts have given new opportunities for boosting the standing of DCEP. As the US dollar continues to weaken due to excessive issuance, as well as the unilateral freezing of US dollar assets and foreign exchange reserves of sanctioned countries by European and American countries, more national concerns have also been strengthened. [6]. As an alternative to the US dollar, the China yuan has entered the vision of financial experts in more and more countries.

The internationalisation of DCEP is also a way to internationalize the RMB. To achieve an internationalized digital economy ecology that recognizes the RMB, China needs to internationalize its digital technology infrastructure and digital technology. Just like the internationalization of different telecommunication technologies such as GSM and MACD in Europe and the US decades ago, the internationalization of CBDC is unworkable without the internationalization of digital technology infrastructure.

DCEP has been validated for theoretical and technical as well as practical feasibility and reliability through three batches of pilot cities. The first batch of pilot cities includes Shenzhen, Suzhou, Xiong'an, Chengdu and the venues of Winter Olympics. The second batch of digital RMB pilots in six regions including Shanghai, Hainan, Changsha, Xi'an, Qingdao, Dalian. Moreover, Tianjin, Chongqing, Guangzhou, Fuzhou, Xiamen, and cities set to host the Asian Games in the Zhejiang province are added in the pilot programme. The common feature of these locations is that the digital infrastructure for technology is well developed, and there are many application scenarios for the digital RMB. Therefore, these places can be used as pilots to explore the advantages and risks of DCEP. In the future, the promotion of digital RMB must first improve the seamless connection between the cloud, terminal, support section, server and user end in various places, which need to match the development of communication technologies such as 5G and gigabit broadband in different places. Then, the experience of the pilot will be replicated and expanded to continuously increase the breadth and depth of the use of DCEP. Once risks arise, they will be restricted to the sandbox of the pilot at different stages of expansion, and problems will be identified and solved in the process of development. Eventually, the nationalisation and internationalisation of the digital RMB can be realised.

According to the most recent PBoC data, as of the end of 2021, there were more than 8 million digital RMB application scenarios, 261 million personal wallets were opened, and the total transaction value was 87.565 billion yuan. These huge pilot data not only fully verified the feasibility of the digital RMB's own circulation, but also provided new guidance for the direction of digital technology development.

With the continuous internationalisation of digital technology in China, the scope of the digital RMB pilot and the process of internationalisation will continue to expand and accelerate. However, it cannot be ignored that a historical problem in the process of RMB internationalisation has not been resolved. The internationalisation of digital RMB must have exchange ports and cash-out channels, etc., which touches on China's foreign exchange controls. According to the regulations, Chinese residents can bring a maximum of USD 5,000 or RMB 20,000 in foreign currency out of China, and a maximum of USD 50,000 or RMB 300,000 can be remitted out of China in a year. If over USD 5,000 is brought in China, a declaration must be made to China Customs, and a "Foreign Exchange Carrying Permit" must be applied if over USD 5,000 is brought out of China. The policy does not allow or encourage residents to engage in foreign exchange speculation, which is obviously in conflict with the internationalisation of the RMB.

A mature financial service system and financial market allows the free exchange and circulation of currency, and even a mature and stable currency derivative trading market to enhance the resilience of the currency ecosystem. Due to special historical reasons such as foreign exchange controls, the RMB cannot become a popular currency in global circulation. The difficulty of RMB internationalization is not only the suppression of competition by international powers, but also the reasons for China's own restrictions. However, the digital RMB is currently in the midst of a period of domain localization, so the issue of internationalization is not as urgent.

### 5. Three Stages of Digital RMB Development

The development of digital RMB generally go through three main stages. The first stage is the infrastructure construction, top-level design, and pilot tests of digital RMB, and China has basically completed the main tasks of the first period. Stage two is the expansion of digital RMB in China's internal circulation and consumption, as well as testing the effect of the central bank's monetary control policy on digital RMB issuance and the impact of digital RMB on the paper money system, and even to use the negative interest rates and other unused monetary tools to enhance regulation [7]. The third stage will focus on the globalisation and internationalisation of digital RMB to match China's national power and economic development level. All three stages will keep pace with the development of China's digital technology. Digital RMB serves and is supported by digital technology. China will improve the digital economy system internally, replicate and expand this model externally, and gradually internationalise the digital RMB.

Likewise, laws and regulations and international agreements and contracts will be synchronized with the development of digital RMB. The widespread circulation of digital RMB will require a reworking of the existing legal provisions on fiat money. In the *Law of the People's Republic of China on the People's Bank of China* and the *Regulation of the People's Republic of China on the Administration of Renminbi*, the scope of fiat currency will be expanded from banknotes to banknotes and digital RMB. Besides, the internationalization of the digital RMB also requires safeguards in the form of multilateral agreements and contracts from partner countries, international cooperation in the form of signing or acceding to treaties, conventions or agreements, and strengthened legal regulation of cross-border payments of the digital RMB. With the weakening of the US dollar dominance, more and more countries are seeking and accepting new settlement methods. In particular, the export pricing, trading and exchange listing of petroleum, new energy, heavy metals, and Chinese specialty commodities are the main application areas for the internationalization of digital RMB. Since 2009, China has signed bilateral local currency swap agreements with a number of countries including Malaysia, Belarus, South Korea, Iceland and Singapore, established the RMB Qualified Foreign Institutional Investor (RQFII) pilot and continuously expanded the quota to more countries. This is a successful experience in the internationalization of RMB banknote, which provides important support and inspiration for the future internationalization of digital RMB.

As of now digital RMB is in the transition stage from the Stage 1 to Stage 2. The promotion of digital RMB requires the support of application channels and scenarios. As a fiat currency, it should be unconditionally accepted within the scope of national sovereignty, and the digital nature also means that digital RMB payment settlement must be unconditionally recognized and supported in mobile payment scenarios. To achieve this, the Chinese government has guided and required mobile payment companies, such as WeChat and Alipay, UnionPay's Cloudflare, Meituan and Pinduoduo, to interoperate between digital currency and paper money at their payment ports. When the digital domain is interconnected, the promotion of digital RMB and everything will come to fruition. However, the transition phase is also fraught with challenges. At present, public acceptance of digital RMB is still very low, and there are many overlapping scenarios for the use of digital RMB and mobile payment, and the hasty introduction of restrictions on overlapping use may do more harm than good.

Under the trend of the promotion of digital RMB, all related parties must grasp the positioning. With the general trend of promoting digital RMB, it is important for all parties to grasp their own positioning. The central bank embarks on the currency issuance and payment business, while private companies can only do wide area payment business or special narrow area environment token business, such as mobile game currencies, which is impossible to change to fiat money theoretically.

## 6. The Promotion and Application of Digital RMB

Money is made for the economy, and digital currencies and the digital economy are inextricably linked. With the explosion of cryptocurrency such as Bitcoin in recent years, an ecosystem of digital economy based on cryptocurrency has emerged. In response to this trend, Facebook, known for its global design network, changed its name to Meta Platforms. Represented by Meta, many companies are looking to shift from a real community platform to building a digital community platform. Platform builders populate digital communities with things that have real-world characteristics, using NFT technology to enable mutual cashing of virtual assets with digital currencies and electronic money. However, the illegal cost of the virtual world platform is low, and the legal guarantee for authenticating ownership is weak. Different platforms can develop similar scenario assets, which are completely incomparable to the uniqueness of real-world resources. In addition, different metaverse platforms are not connected. The grabbing of stock customers and the safety and security of their own platforms are challenges. Companies as managers and designers of metaverse, are prone to lose fairness and legal security. This is the original problem that NFT needs to solve.

However, it should be pointed out that digital currency includes digital fiat currency and private digital currency. Although both are named with digital concepts, their connotations are different. The private digital currency, take Bitcoin as an example, is an asset-based currency that originated in a niche, or even used in a niche, and because of its better encryption and natural limitation, it has been identified by a particular industry chain as a sort of asset hedge and a way for assets to evade regulation. And for the majority of people, Bitcoin is just a tool used for speculation, not for payment and settlement. At the same time, because of its scarcity, its inherent deflation belongs to the properties of metal-like money and cannot become legal tender under the credit system, because fiat money theoretically needs to be issued indefinitely to meet the infinite growth of economic development.

The digital RMB's digital technology makes it a tool for measuring the value of data assets and serves as a medium of exchange for trading data assets, realizing the direct connection between currency and data assets, which is better than private digital currencies. [9]

Digital RMB is an innovation of digital currency. It can be circulated without barriers in the traditional economy and digital economy under the guarantee of legal system, which can greatly promote the development and integration of traditional economy and digital economy. These aspects cannot be done by private digital currencies such as Bitcoin, nor by other electronic currencies.

Private digital currencies are essentially in competition with banks based on a credit system. However, digital currency currently is speculated as a currency with value-added, and only plays the attribute of metal-like currency, while its function in the digital economy is weakened. It is believed that in the future, digital currencies will inevitably move towards the bank profitability of deposit and loan earning spread, and play its monetary attributes.

For commercial banks, digital currency is both a challenge and an opportunity. Traditional commercial banks must accelerate the transformation to digital banking and improve the upgrade for digital currency transactions, storage, lending and other functions as soon as possible. The digital RMB issuance rules show the central bank's precise real-name and flat management of the currency, which is a huge advantage in macro-regulation and micro-tracking. However, for commercial banks under the current credit system, it is undoubtedly a big negative, which will compress the profit methods of commercial banks, reduce the money multiplier effect, further reduce the volume of money supply M0, and even reduce the volume of M1 and M2. These take away commercial bank profits disguised as financial inclusion, lowering the cost and threshold of market financing, which is good for the public. Many people believe that digital RMB issuance, is increasing M0, however, in the absence of the money multiplier effect and commercial banks' storage and lending amplification effect, it is shrinking the money supply, including M0, M1 and M2. Therefore, the digital RMB has continued to be adjusted in the banking sector of China's A-shares from the time it was set. The transformation of the banking industry from traditional banks to digital and technology banks has become imminent and trendy.

Current direct competitors to the digital RMB are private digital currencies and potential competitors are digital fiat currencies of other countries. The rollout phase of the digital RMB is also bound to see the emergence of a business model based on the spread between deposit and lending of currency attributes. Currently, digital RMB is positioned as M0 only at the design stage, but many commercial banks have set up mechanisms for cashing out bills in digital RMB in order to enhance people's recognition and acceptance of digital fiat money, which conflict with the digital RMB's position. The reason is that, after the digital RMB is cashed into paper money, it may be used for wealth management, savings or even private lending, rather than just consumption. Merchants usually use the bills for investment and savings for interest purposes. Customers who convert digital currency into paper money may be more inclined to use it in investment and savings rather than consumption. The only way to avoid this is to set interest rate differentials and increase the cost of encashment. But the reality is that the digital RMB is not a necessity right now, and once the cash in cost is set, it is not conducive to the implementation of the promotion. Furthermore, when larger amounts of digital RMB are directly transferred from M0 to M1 and M2, it can cause deflation and disruption of central monetary policy [10]. In conclusion, accurate placement of digital RMB can be done, but accurate spending is more difficult, especially if it can be cashed out, which will break the definition of the M0 attribute of digital RMB and trigger financial risks.

## 7. Digital Economy

The digital RMB is not only based on Beijing-Tianjin-Hebei, Yangtze River Delta, Guangdong-Hong Kong-Macao Greater Bay Area and international trade along the Belt and Road, but also constitutes one of the necessary elements of digital economy system in China to provide new growth and enhance the efficiency of China's economic development. [11]

President Xi Jinping emphasized that the construction of a "dual circulation" development pattern in which the domestic economic cycle takes the lead while the international economic cycle serves as an extension and supplement is a direct reflection of productive force reform and development. The new economic development pattern is dominated by the domestic economic cycle, and take the advancement of the supply-side structural reform as the main thread, and focuses on the digital construction and interoperability of production, distribution, circulation and consumption in all aspects. In the production side, the government introduces high-end production factors and scarce resources to make up for domestic production needs and improve digital hardware and technical foundations. In terms of distribution, the government has strengthened the infrastructure to realize smart scenes, smart villages, smart cities and smart countries, increasing employment and income. On the circulation side, the government is working to improve the efficiency of the industry and solve problems such as obstacles to internal flows. On the consumption side, the government is committed to delivering high-quality goods to match people's aspirations for a better living. [12]

In April 2020, the General Office of the State Council issues a notice about *Comprehensive Reform of the Market-oriented Allocation of Factors*, which for the first-time listed data, land, labour, capital, technology as the five elements. It directly provides the theoretical foundation for data, the core element of the digital economy. [13]

In 2019, the added value of the digital economy in 47 countries around the world reached US$31.8 trillion, and the digital economy accounted for 41.5% of GDP. The average growth rate of the global digital economy reached 5.4%. The total output value of China's digital economy reached 35.8 trillion yuan, accounting for 36.2% of GDP. The total digital economy increased by 12.7 times compared with 2005, and the growth rate was as high as 20.6%, much higher than the average growth rate of the global digital economy during this period. [14]

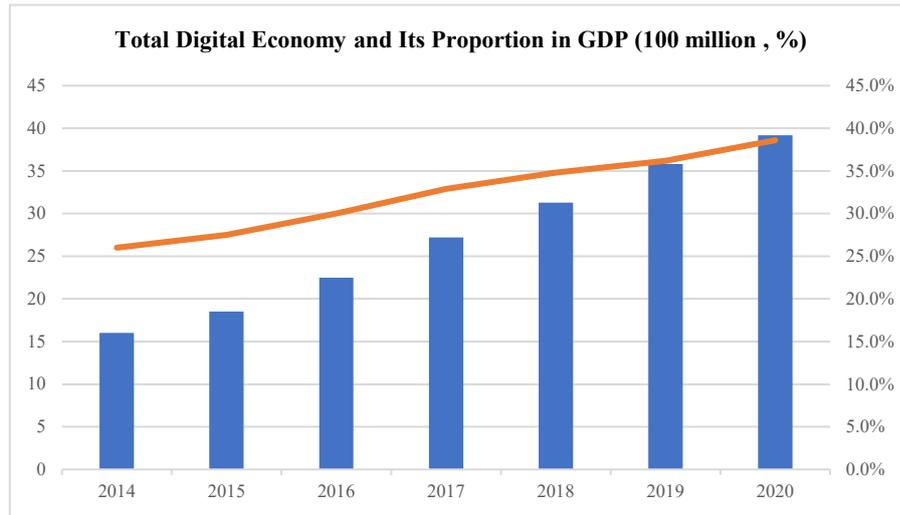

Figure 2: China's Total Digital Economy and Its Proportion in GDP

China's digital economy is based on the foundation of the Internet economy since 2000 and has already played its role in the "dual circulation" development pattern. Issued in December 2021, the "14th Five-Year" Development Plan for The Digital Economy defines the digital economy as the direction of economic transformation and a new growth point. The details are shown in the Table 1.

| Indicators | Year 2020 | Year 2025 |
|---|---|---|
| The added value of the core industries of the digital economy as a percentage of GDP (%) | 7.8 | 10 |
| Number of active IPv6 users (100 million) | 4.6 | 8 |
| Number of gigabit broadband users (million) | 6.4 | 60 |
| Scale of software and information technology service (trillion) | 8.16 | 14 |
| Application penetration rate of industrial Internet platform (%) | 14.7 | 45 |
| Online retail sales (trillion) | 11.76 | 17 |
| E-commerce transaction scale (trillion) | 37.21 | 46 |
| Scale of e-government services users (100 million) | 4 | 8 |

Table 1: Main Indicators of Digital Economy Development During The 14th Five-Year Plan

The digital economy is an ecosystem that is highly dependent on digital technology in terms of functional flow and information interaction. It is an upgraded form of the traditional non-digital economy model. Except for a few special scenarios, the conventional perception of the digital economy can be summarized in Figure 3.

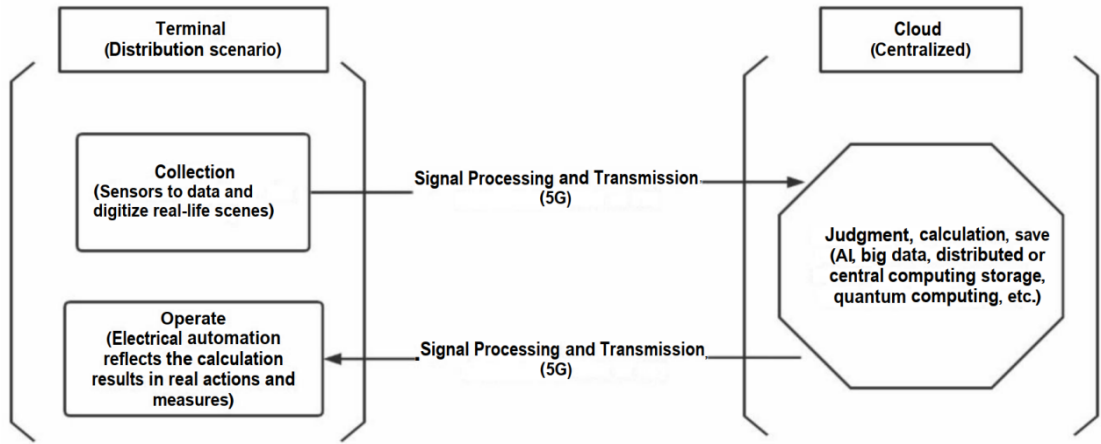

Figure 3: Digital Economy Operation Framework

The digital economy is widely involved in AI, cloud computing, big data, and technologies such as digital signal processing and transmission, and electrical automation. This complete industrial value chain, from data generation, acquisition, transmission, to modelling, computing, processing, storage, etc., and then the results of the signal transmission back to the operator, through electrical automation, robotics, and other scenarios, to achieve the value of data embodiment. As the supervisor or beneficiary, citizens will get efficient and accurate service and product experience in this process.

Combined with the current national conditions, the development of the digital economy will focus on the following aspects.

(1) Digital Government Construction
Through online government service platforms, residents can apply for government services online. It is expected that online government services will improve the transparency of government affairs and even the openness of data packages, which will give rise to the flourishing of new industries such as assistance services and data analysis. Digital RMB and digital ID cards will gradually become a necessity in people's lives, and there may be a strong binding requirement for both. Some digital platforms in villages, with TV screens and cell phone screens as terminals, have realized services and information to reach the people precisely, realizing the sinking of quality resources such as medical care, education, and agricultural technology, allowing farmers to truly enjoy the benefits of modern development with digital forms of resources.

At the same time, the combination of digital government platform and digital RMB is a kind of accurate supervision for the local government. From project budget to execution, the flow and use of funds can be accurately monitored through digital RMB appropriation and debt release, preventing government employee's corruption and lazy and neglectful of duties. Eventually, the performance of government can be objectively judged by the proportion of input to output of digital RMB.

(2) Digital Countryside Development

E-commerce in digital villages, will benefit from the sinking of digital economy technology into the rural field, providing new growth points for the economic development of the vast number of towns. Digital RMB will have a huge positive effect on financial inclusion. China's rural Internet users have reached more than 284 million, and the national rural online retail sales amounted to approximately 2.05 trillion yuan, and the agricultural products online retail sales amounted to more than 422.1 billion yuan. According to *Action Plan for the Development of Digital Countryside (2022-2025)*, by 2025, important progress will be made in the development of digital countryside, and a number of rural e-commerce product brands with high visibility, good quality and characteristics, rural culture and rural digital governance system will be increasingly improved.

During the development of rural digital economy, commercial banks will use digital RMB to give farmers a new inclusive policy of financing and lending. Moreover, the popularity of digital RMB in the cyberstar economy will serve farmers more conveniently and safely, and combat tax evasion and false sales.

(3) Technology Companies Development

The Internet industry and software and information services industry are expected to flourish with the development of the digital economy. The digital RMB will play a great role in businesses for transaction and financing, stimulating the development of innovative technologies such as AI, cloud computing, big data and blockchain. Besides, the digital RMB will facilitate entrepreneurs to make meticulous observation and judgment on industry development and market segmentation. This stimulates the "Entrepreneurship and Innovation among All the People" policy to a certain extent.

(4) Transaction Supervision

For the cross-border transaction, the traditional cross-border transactions usually use corporate accounts to settle payments in a uniform manner with monthly, quarterly or even annual bookkeeping. However, the payment process sometimes may result in commercial disputes over misplaced payment numbers, and can even become a channel for criminals to launder money. The point-to-point accurate payment function and double offline settlement feature of digital RMB greatly facilitate cross-border real-time transactions. Controlled anonymity also enables accurate monitoring of illegal money laundering practices while safeguarding the privacy of both parties to the transaction.

Furthermore, with the completion of the overall layout for the national integrated big-data centre system, a large number of digital business scenarios, such as data processing, transportation, computing and other upstream and downstream business interactions in various segments, will generate demand for digital RMB settlement. The government will set up a professional data exchange. Existing stock, futures, and equity exchanges may pilot the requirement of opening accounts with digital RMB as a tied account. In the future, the account opening rules will be upgraded to require all accounts to be tied to digital RMB for transactions and operations, replacing the current rules for tied bank card accounts, which has the risk of customer information leakage and money laundering crimes.

The number of 5G and gigabit broadband users in Chins are expected to exceed 600 million by 2025, enabling a seamless connection between a large number of terminals and the cloud and laying a complete and high-speed infrastructure for digital RMB promotion. As the digital economy matures, digital trading will usher in a phase of rapid development, with various exchanges emerging and even digital trading intermediary services following suit.

**8. Application of Digital RMB in Digital Economy**

Money arises from the economy and serves the economy, and so does digital currency. Digital currency serves the digital economy, which is mainly apply in business transactions and personal consumption. Expanding the application scenario of digital currency is the current focus of promoting digital RMB, which can be started from the following directions.

First of all, the green finance is the area where the digital RMB can make a big impact. "The PBoC will continue to explore the use of financial technology to develop green finance. The application of big data, artificial intelligence, blockchain and other financial technology tools in green finance holds great promise." said by Yi Gang, the governor of PBoC. The size of China's domestic and foreign currency green loans exceeded 13 trillion yuan by 2021, accounting for 7.18% of the total domestic credit stock of 181 trillion yuan. (Figure 4) Additionally, the size of green bond stock exceeded 1 trillion yuan, ranking second in the world. Policies will strongly support the green finance sector. Green financial products such as green credit, green bonds, green leases and green trusts can only be effective when the funds are used for green purposes. In August 2020, China UnionPay launched a digital bank card in conjunction with 12 banks. It is predicted that a digital credit card dedicated to green industry finance may also be launched in the future [15].

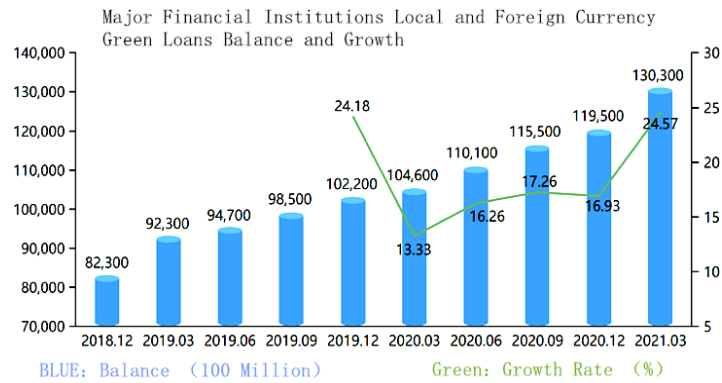

Figure 4: Balance and Growth Rate of Domestic And Foreign Currency on Green Loans of Major Financial Institutions. Source: Wind

Secondly, foreign trade, especially commodities, will also be the main area for digital currency. As the international and domestic environment is changing day by day, and the commodity trading market has been gradually seeking settlement in currencies other than the US dollar. China, as a major buyer of commodities, can gradually use approach that benefits itself by establishing international contracts and gradually promoting digital RMB settlement options starting with friendly countries and Belt and Road countries.

Thirdly, consumer areas such as digital payment scenarios combined with digital ID cards will also benefit from the promotion of digital RMB. More and more scenarios for identification and binding digital economy appear, and digital ID cards have come into being. The combined use of digital ID cards and digital RMB can conveniently serve the payment fields with high identification requirements.

In addition, the digital RMB can use technology to enable the regulation of digital government and industry rules. Digital currency has inherent advantages in finance. For example, digital RMB can achieve targeted and fixed interest rate differential treatment in financing loans; for bonds, it can achieve traceability in the use of funds, enhance default risk mechanisms and return mechanisms and avoid illegal acts such as money laundering and pyramid schemes. According to statistics in 2018, China recovered 3.54 billion yuan of stolen money in cases of combating job-related crimes, investigated and dealt with pyramid schemes involving 8.72 billion yuan, and cracked smuggling cases involving a value of nearly 7 billion yuan. If the traceability of digital RMB's blockchain technology is utilized, both regulation and retrieval of funds can be turned into a cloud business and automated, greatly reducing the cost to state public officials.

Moreover, controlled anonymity and block-tracking technologies make financial regulation of digital RMBs extremely easy, while the cost of breaking the law becomes extremely high, which can effectively curb illegal financial activities. Currently, the Chinese tax system has also proven the feasibility of big data for taxpayer review and tracking. Both the vast data flow and the huge amount of money flow can be accurately measured down to the person.

At the same time, digital RMB and digital ID cards, digital enterprise licenses and other licenses are likely to be strongly bound, while weakly bound to face recognition and fingerprint recognition of cell phones to realize digital data generation and supervision and management of real people's identity and asset trajectories, etc. These data and digital currency, digital transactions, digital transmission, digitization, digital arithmetic will constitute the elemental part of the digital economic system.

## 9. Summary

After the Russia-Ukraine war, economic globalisation turned into economic semi-globalisation. In the process of de-globalisation, the games in economic development began to shift from the world scale to the regional scale. Stagflation has become a fact under the environment combined with the current currency overdraft, the impact of the epidemic and various factors such as the energy crisis and war sanctions. With a large time horizon, a wide range of countries and industries affected, and few ways to mitigate the problem, this stagflation may well be the most difficult stagflation in human history to deal with. The recession that will immediately follow the stagflation could also be the biggest in human economic history. Perhaps this is a bit pessimistic but looking at the typical stagflation in human economic history, the triggering factors, structural factors and ideological confrontation are indeed the most serious since the Second World War. Moreover, we seem to have abandoned the most useful tool for alleviating stagflation - using technological innovation and economic globalisation to reduce costs and stimulate consumption.

The confrontation between the East and West camps has reached its peak, and henceforth technological innovation based on people's welfare will give way to technological innovation based on national confrontation in the military race. And the result of de-globalisation will inevitably lead to waste of resources, reduced productivity efficiency, duplication of R&D and manufacturing, etc. Combined with the decline in consumer demand and the decline in production and investment enthusiasm, a vicious cycle of economic recession seems to have become the gray rhino in front of the world.

For Russia in 2022, its dollar and euro foreign exchange is extremely insecure, and this incident has also prompted other countries to seek a new world currency position, and even Saudi Arabia has made it clear that it will use the renminbi as the currency for pricing and trading oil. These major international events have provided opportunities and challenges for the digital renminbi.

To get out of the rut and out of stagnation, the digital economy, an efficient economic model, and digital currency, a new form of fiat money, have become the best choice.

Economic model progress and technological consumer upgrades have been the source of every progress in history. In this historical change, which country steps on the right rhythm and is at the forefront, which country will be the biggest beneficiary.

Since the 14th Five-Year Plan, China has been simultaneously building digital infrastructure and implementing digital scenarios in the manufacturing power, data exchange construction, high-speed network transmission and quantum computing, etc. The mutual promotion of digital RMB and digital economy has entered a high-speed stage. Internationally, the sanctions on Russia's foreign exchange assets have also triggered potential demand from many countries seeking new foreign exchange currencies. All of these new opportunities trigger the development of China's digital RMB and digital economy, and are well worthy of active reference and study by countries around the world.

Reference

[1] Wang Yongli. "The Essence and Context of "Digital Currency". Published in The Economic Observer on August 31, 2020.
[2] Qi Yudong, Liu Huanhuan, Xiao Xu. Digital Currency and International Monetary System Reform and New Opportunities for RMB Internationalization [J]. Journal of Wuhan University (Philosophy and Social Sciences Edition), 2021(5).
[3] Ji Xiaonan, Chen Shan. Discussion on the mechanism and countermeasures of legal digital currency affecting RMB internationalization [J]. Theoretical Discussion, 2021(1).
[4] Jiang Chen. Challenges faced by the central bank in issuing digital RMB and corresponding countermeasures [J]. Science and Technology Economic Market, 2021(2).
[5] Zhang Shuzhe. The system architecture of digital RMB and the impact of its issuance on economic operation [J]. Enterprise Economics, 2020(12).
[6] Wu Yunyun. The legal digital currency of the central bank and the internationalization of RMB [J]. Modern Business, 2021(3).
[7]Huang Guoping, Ding Yi, Li Wanrong. Development Trend, Impact and Policy Suggestions of Digital RMB[J]. Research on Financial Issues, 2021(6).
[8]Qiu Yanfei. Obstacles and Legal Paths of Digital RMB to Realize Cross-border Payment [J]. Finance and Economics, 2021(11).
[9] Liu Dian. Digital RMB: Ecological Reconstruction and Global Competition of Digital Economy [J]. Cultural Landscape, 2021(1).
[10] Li Wenjuan, Zhang Yinghua. Analysis of the operation mechanism of digital RMB [J]. National Circulation Economy, 2021(3).
[11]Wu Zhennan, Zhang Xuanming. The development path of legal digital currency under the background of digital economy [J]. International Business Accounting, 2021(10).
[13] Wang Wentao. Promoting the Construction of a New Development Pattern with a High Level of Opening-up"

[14] Liu Xiaoxin. Analysis of the main characteristics and impact of digital RMB [J]. People's Forum, 2020(26).

[15] China Academy of Information and Communications Technology: White Paper on China's Digital Economy Development (2020), July 2020, p. 10.